\documentclass[fleqn,twoside]{article}
\usepackage[headings]{espcrc2}

\readRCS
$Id: espcrc2.tex,v 1.2 2004/02/24 11:22:11 spepping Exp $
\usepackage{graphicx}
\usepackage[figuresright]{rotating}

\newcommand{\AmS}{{\protect\the\textfont2
  A\kern-.1667em\lower.5ex\hbox{M}\kern-.125emS}}

\title{IceCube: Performance, Status, and Future}

\author{Carsten Rott\address[PSU]{Physics Department, Pennsylvania State University,
        104 Davey Lab,\\ 
        University Park, PA 16802-6300, USA} for the IceCube Collaboration\thanks{http://icecube.wisc.edu}\\
http://icecube.wisc.edu/science/publications/isvhecri2006.html
               }
       
\runtitle{IceCube: Performance, Status, and Future}
\runauthor{C. Rott}

\begin{document}

\begin{abstract}
High-energy neutrinos are uniquely suited to study a large variety of
physics as they traverse the universe almost untouched,  
in contrast to conventional astronomical messengers like photons or
cosmic rays which are limited by interactions with radiation and matter at high
energies or deflected by ambient magnetic fields. 
Located at the South Pole, IceCube combined with its predecessor
AMANDA comprise the world's largest neutrino telescope. IceCube
currently consists of nine strings, each containing 60 digital optical
modules, deployed at depths of $1.5$ to $2.5$~km in the ice and an array of 16
surface air-shower stations. IceCube is expected to be completed in early 2011
at which time it will instrument a volume of one km$^3$ below
the IceTop air-shower array covering an area of one km$^2$.
The current IceCube detector performance is described and an
outlook given into the large variety of physics that it can address, with
an emphasis on the search for ultra-high-energy neutrinos which
may shed light on the origins of the highest energy cosmic rays.

\vspace{1pc}
\end{abstract}

\maketitle

\section{Introduction}

Neutrinos open up a new window to the cosmos as they propagate undisturbed through the universe and
therefore allow an undistorted view in contrast to messengers like photons or protons.
Electrically charged particles bend in magnetic fields, therefore for protons below 10~EeV ($10 \times 10^{18}$~eV)
any directionality information is distorted. 
Furthermore, above 50~EeV protons 
interact with the cosmic microwave background radiation (CMB). This is known as the
Greisen-Zatsepin-Kuzmin (GZK) cut-off~\cite{GZK}.
Neutrinos, on the other hand, allow one to study the entire energy spectrum.

\section{IceCube}

IceCube will instrument a volume of approximately one cubic kilometer with at least 70 strings arranged in a hexagonal pattern
of $125$~m string spacing. Each string contains 60 Digital Optical Modules (DOM) spaced 
evenly between a depth of $1450$~m to $2450$~m.
Each DOM consists of a $32.5$~cm diameter pressure sphere 
that holds a downward facing $10"$ 
diameter HAMAMATSU R7081-02
photomultiplier tube (PMT) with ten dynodes supported by coupling gel, a signal processing electronics board, an LED flasher board 
for calibration, and a high voltage base which powers the PMT at a gain of $ 10^7$. 
Signals above a threshold of 0.2 photo electrons (pe) are digitized in the ice by both a
Fast Analog-to-Digital Converter (FADC) that 
samples at a fixed rate of $40$~MHz for $6.4$~$\mu$s and
an Advanced Transient Waveform Digitizer (ATWD) sampling 128 bins of $3.3$~ns width.
To extend the dynamic range, the ATWD contains three channels with gains of $0.25$, $2$, and $16$.
IceCube sensors have a substantially lower noise rate, while having an optical sensitivity that is about $1.4$ times
higher compared to its AMANDA counterparts.
The dark noise rates of the DOMs in ice (with after-pulse suppression of $51~\mu$s) 
is about $700$~Hz ($350$~Hz).
This rate is further reduced to about $10$ to $20$~Hz by a local coincidence logic which 
requires activity in at least one of the adjacent DOMs in a $2~\mu$s time window.
Above each IceCube string there is one IceTop surface station that consists of two tanks, each 
holding a low ($5\times 10^5$) and a high gain ($5\times 10^6$) DOM. This IceTop air shower array~\cite{Bai} is an 
integral part to the IceCube detector and relies on the same technology.

Before DOMs are deployed, they have to pass a rigorous test procedure. 
In addition, for a few DOMs a complete study of the absolute quantum efficiency and
collection efficiency variations over the surface was conducted~\cite{Miyamoto:2005qr}, 
to help determine systematic uncertainties related to the DOM surface efficiency variations, which 
was a limiting factor in many AMANDA analyses.


\section{Construction}

The IceCube detector is constructed at the South Pole during a short construction period 
lasting from early December to late January.
During the first season 2004-2005 one string and four IceTop stations were installed, followed by eight more strings and
12 IceTop stations in 2005-2006.
String deployment proceeds in less than $12$ hours, into the $60$~cm diameter holes, previously melted by 
an enhanced hot water drill to a depth of $2500$~m in less than 35 hours.  
Information about ice properties was gathered through an {\it in situ} measurement with 
a system called the dust logger, which is lowered into the drill hole. 
This device sends out  $410$~nm laser pulses at $120$~Hz.
Brushes, blocking the drill hole, ensure that light has to propagate through the ice, before 
being collected at the bottom of the system. 
In this way, the scattering and absorption coefficient have been determined with very high precision~\cite{dustlogger}.

\section{Detector Performance}


Beginning with the deployment of the first string on January 28, 2005, IceCube's performance has been studied
intensively. The first year single string data were found to be consistent with expectations~\cite{performance}.
IceCube's current multistring configuration allows for a more complex performance evaluation, which 
is the focus of this review. 

The detector performance can be evaluated using data itself, complemented with simulation
studies to understand uncertainties in the analyses better.
Down-going muons produced naturally in cosmic ray showers transverse the detector and thus provide an ideal sample
to study the detector. 
In addition, IceCube has artificial {\it in situ} light sources - 12 separately flashable LEDs for each DOM -
that can be seen by adjacent DOMs and are ideal for studying a variety of detector properties. 

The down-going muon sample is used to test the track reconstruction
algorithms and also to study detector performance. 
They are ideal for long term stability studies, as they are
continuously present in the physics data stream.
Muon tracks are reconstructed with the standard IceCube offline reconstruction software that
uses a log likelihood fit. Full waveform based reconstruction algorithms are currently
under development. 

\subsection{Timing}

Any physics analysis in IceCube relies on the precise knowledge of the relative timing of all the DOMs 
throughout the whole detector.
We studied the timing variations of individual modules and
the consistency of the timing of DOMs with respect to each other and found them to be consistent with
design requirements.

\subsubsection{Timing Resolution}
\begin{figure}[htbp]
  \begin{center}
    \includegraphics[width=3.5cm]{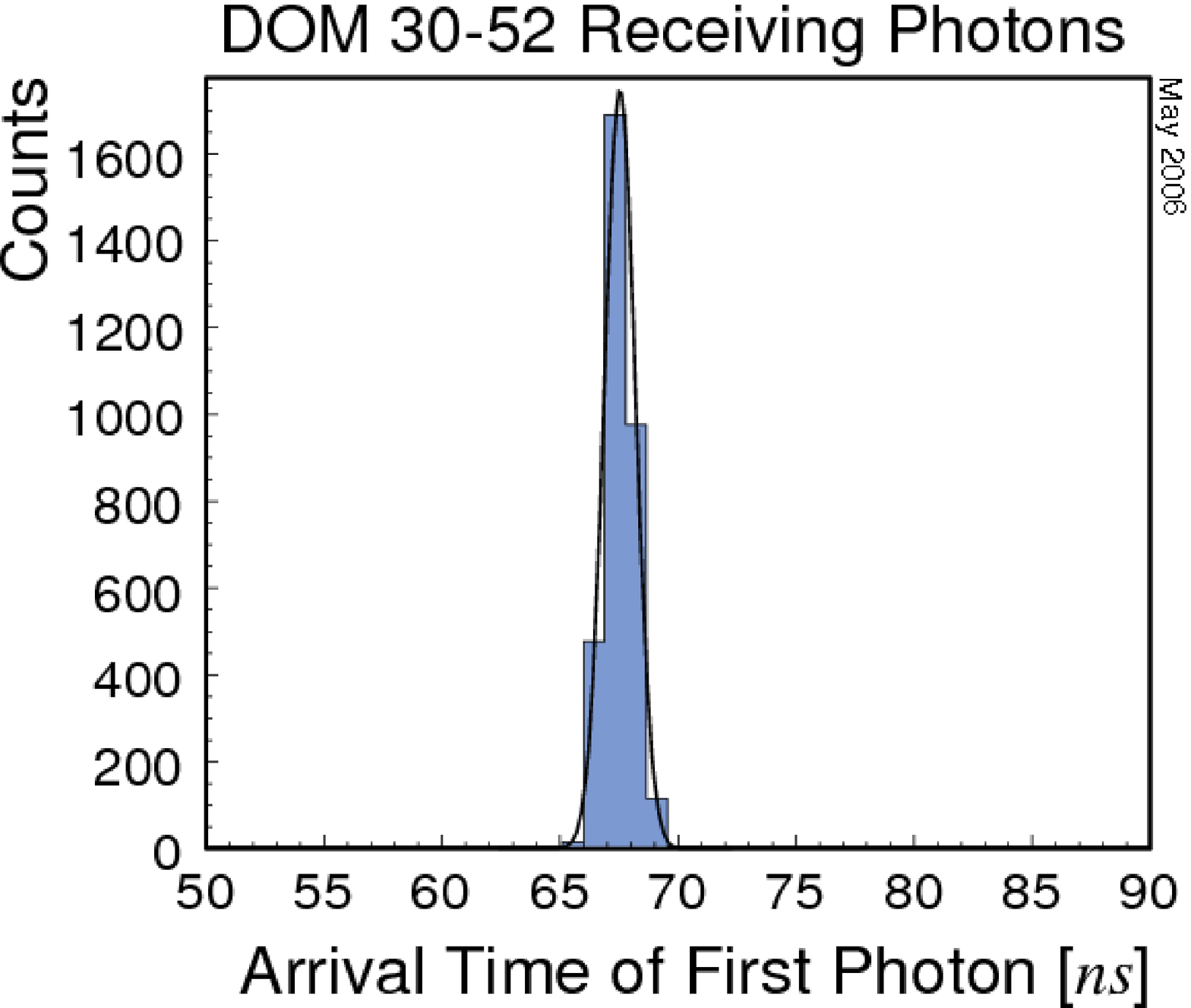}
    \includegraphics[width=3.5cm]{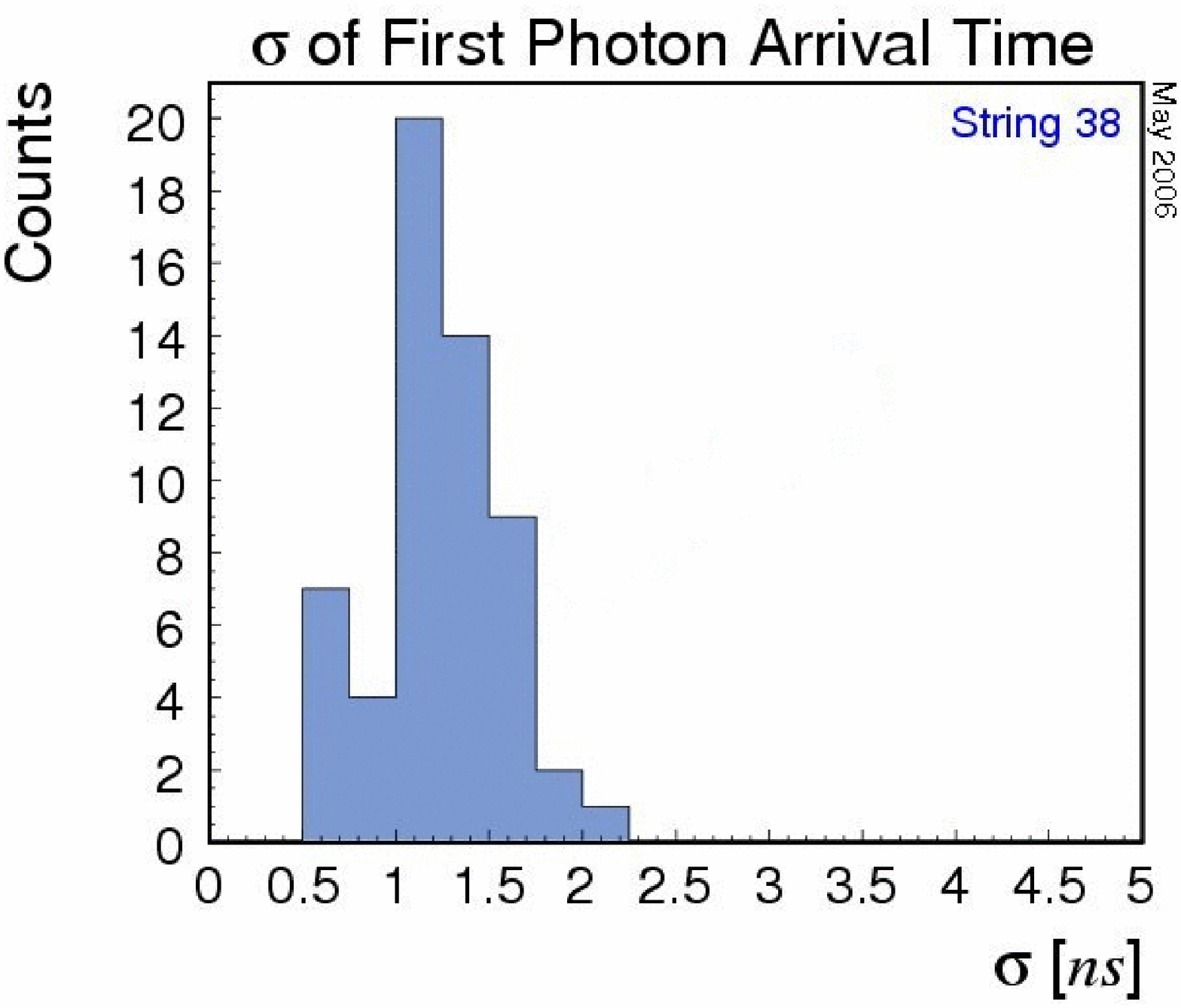}
    \caption{Left: The earliest hit time distributions from flasher data along with the fit to a Gaussian for a 
                   representative DOM. 
	  Right: The variation in hit time for the DOMs on a string.\label{flash_hittime}}
  \end{center}
\end{figure}

The timing resolution of the DOMs has been studied in flasher data.
For this, we determined the time difference between a flashing DOM and the time of the first hit on the 
DOM above it. The variation in the earliest hit time is related to the ice properties and the intrinsic timing resolution 
of the observing DOM. It is found to be better than $2$~ns to $3$~ns (see Figure~\ref{flash_hittime}).
This result was also verified with several independent methods, using calibration
signals (Rapcal)~\cite{performance}, flashers, and down-going muon data.

\subsubsection{Timing Consistency}

To check for consistency of the timing of an individual DOM with respect to the entire detector,
we calculate the time residual, defined as the difference between the expected Cherenkov light arrival time 
from a reconstructed track and the observed hit time in a given DOM.
The DOM under study is removed from the fit to avoid biasing the reconstruction.

\begin{figure}[htbp]
  \begin{center}
    \includegraphics[width=7cm]{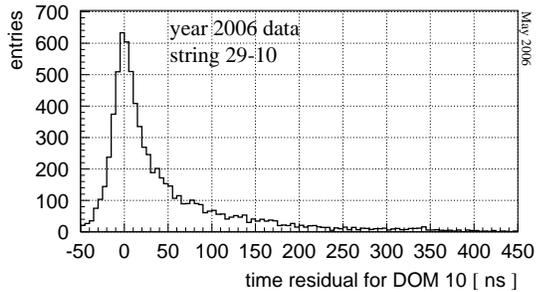}
    \caption{Time residual distribution from fits to down-going muons for a typical DOM.\label{time_residual}}
  \end{center}
\end{figure}

\begin{figure}[htbp]
  \begin{center}
    \includegraphics[width=7cm]{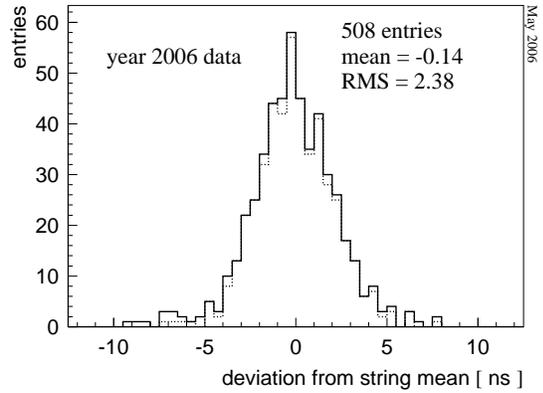}
    \caption{Deviation of the prompt peak position of the time residual distributions for ``in ice'' DOMs from the string mean.\label{TRH_prompt_peak}}
  \end{center}
\end{figure}

The time residual distribution (see Figure~\ref{time_residual}) is expected to be peaked at zero if
the timing of the module is consistent with the detector or otherwise shifted.
The positive tail in the time residual distribution is from light that scatters 
before arriving at the sensor. 
Figure \ref{TRH_prompt_peak} shows the obtained prompt peak positions of the individual DOMs. The distribution
is consistent with zero, likewise the spread arising predominantly from uncertainties in track reconstruction
is within expectations.

\subsubsection{Long-term Timing Stability}

\begin{figure}[htbp]
  \begin{center}
    \includegraphics[width=7cm]{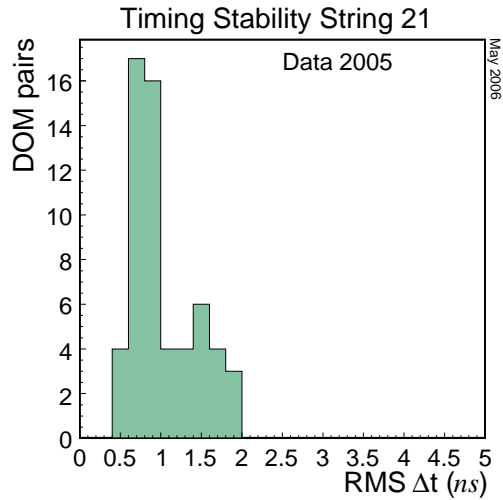}
    \caption{Variation of time difference between adjacent DOMs for selected, nearly vertical
             down-going muon events over the course of 2005.\label{dt_rms_str21}}
  \end{center}
\end{figure}

To verify long-term timing stability within a string, the hit time difference between adjacent 
DOMs in events with a down-going muon running nearly parallel along the string were selected.
The mean hit time difference was determined as a function of time.
Figure~\ref{dt_rms_str21} shows the obtained 
variation between adjacent DOMs over the course of a year. It was determined to be smaller than $2$~ns,
which was also confirmed by determining the variation in the prompt peak position of the
time residuals over the year.

\subsection{Geometry Verification}

The detector geometry was initially determined through a laser range and depth sensor measurements
at the time of deployment. 
The geometry within a string was verified by comparing hit times of adjacent DOMs from down-going muons and
from flashers. The position was verified within about a meter -
consistent with the expected error in the method.
Relative depth of the strings with respect to each other and their distances in the ice were also
verified using flasher data.
In this method, the arrival time of unscattered light from a flashing DOM was determined for observing DOMs on a neighboring 
string by fitting the leading edge of the photon arrival time distribution.
Using the speed of light in ice, the distance between the flashing DOM and each of the observing DOMs was 
obtained from the measured light travel time of earliest arriving photons.
Results were obtained using different string pairs. With this method the geometry information can be verified
within 1.0~m.

\subsection{DOM Occupancy and Efficiency}

\begin{figure}[htb]
  \begin{center}
    \includegraphics[width=7cm]{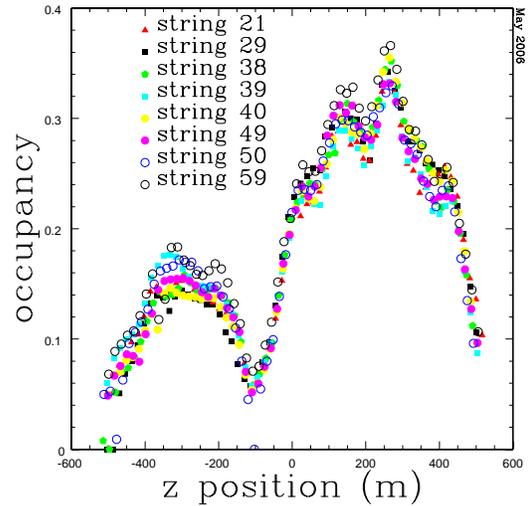}
    \caption{The DOM occupancy in the different strings are consistent with each other and features are clearly 
            correlated with measured ice properties. The expected decrease in muon flux with depth is also observable (z=0 marks
the detector center).\label{dawn_occupancy}}
  \end{center}
\end{figure}

The occupancy for each DOM was determined in the down-going muon sample and 
the resulting distribution normalized by the total number of events for the 
string that satisfied the selection conditions (see Figure~\ref{dawn_occupancy}). 
The occupancy distributions of the various strings are consistent with each other and the observed structure is
correlated with the optical properties of the ice as measured independently~\cite{dustlogger}.

\begin{figure}[htbp]
  \begin{center}
    \includegraphics[width=7cm]{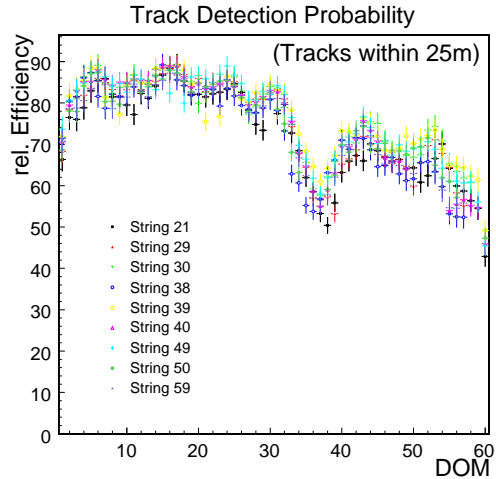}
    \caption{Probability of the DOMs in the ice to observe a hit if a reconstructed down-going muon 
	passed within a given distance.\label{tdp}}
  \end{center}
\end{figure}

The efficiency of each DOM in the ice was studied by using reconstructed
down-going muon tracks to determine the probability that a module observed a hit when a track passed 
within a given distance to it. 
This efficiency is found to be consistent within the strings as shown in Figure~\ref{tdp}.

\subsection{IceTop - InIce Timing }

\begin{figure}[htbp]
  \begin{center}
    \includegraphics[width=7cm]{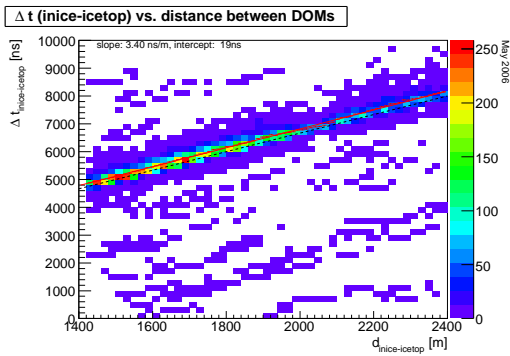}
    \caption{In-ice and IceTop timing is consistent.\label{inice-icetop}}
  \end{center}
\end{figure}

The timing between the IceTop air shower array and the in ice detector was verified by
studying the time and spatial difference between hit modules of the two subdetectors.
Figure \ref{inice-icetop} shows the obtained result, that is consistent with the speed 
of light as expected.

\subsection{AMANDA - IceCube Coincidence}

With its smaller OM spacings and interstring distances, AMANDA provides a compact 
subdetector array within IceCube. 
It is especially interesting for low energy analyses, like searches
for WIMPs or slow moving particles.
The joint operation of AMANDA and IceCube was tested by flashing DOMs in IceCube and observing the
light in AMANDA and by extrapolating muon tracks from one detector into the other.
Both approaches confirm that AMANDA and IceCube can be jointly used and the first combined
analyses are underway.

\section{Physics outlook}

IceCube can address a large variety of physics: Searches for weakly interacting massive 
particles (WIMPS), point sources, diffuse neutrino fluxes, transient sources, supernovae, GRBs, {\it etc}. 
IceCube may also observe spectacular signatures from tau neutrinos which would be very likely 
of extraterrestrial origin.
IceCube's design has been optimized for TeV to PeV neutrino energies but can also
be used to search for neutrinos at higher energies, even though such events might not be 
fully contained within the detector.

\subsection{Tau Neutrino}

The search for tau neutrinos is especially interesting as they are almost certainly of extra-terrestrial origin
and exhibit very distinctive signatures.
Among those signatures is the double bang~\cite{double_bang}: Two cascades are produced inside
the detector volume as a result of a charge current interaction of an incoming $\nu_{\tau}$ and
the sequential decay of the produced tau. At tau neutrino energies $E_{\nu_{\tau}}$ between $10^{15}$ and $10^{16}$~eV, the
cascades are separated enough so that this signature can be observed with IceCube. At higher energies,
the lifetime of the tau becomes large and it is likely that events are not fully contained 
within the detector volume, {\it i.e.} only a single cascade and the corresponding tau track would be observed. 
This type of event is referred to as a lollipop~\cite{Beacom:2003nh}.
Another class of events labeled sugardaddy~\cite{DeYoung:2006fg}, where the tau decays
into a muon resulting in dim tau track followed by a brighter muon track, are also studied.
The uniqueness of the tau signatures, the fact that they are almost certainly of extra-terrestrial origin and that
these signatures have not been observed, makes these events especially interesting. 

\subsection{Extremely High Energy Neutrinos}

Extremely energetic protons will lose their energy via a resonant 
scattering process with a photon in the $2.7$~K CMB radiation, this is the
GZK cut-off: $ p + \nu_{2.7k} \rightarrow \Delta^* \rightarrow N + \pi$.
GZK neutrinos are produced through the decay of the pions.
The detection of these GZK neutrinos, of which IceCube is expected to observe
one event per year~\cite{aya}, would supply firm evidence that the EHE cosmic rays 
are coming from extragalactic space.
Through the study of the neutrino spectrum at GZK energies, we are also sensitive to
annihilation or collapse of topological defects, that could generate EHE 
neutrinos with energies beyond the GZK scale. 

\section{Future}

To obtain an EHE spectrum and perform point source analyses, a large statistical sample of EHE events is 
needed. In order to get a sufficient event rate, a detector of about 100 cubic km would be needed.
The large absorption length for sound and radiowaves in the ice would make it feasible 
to instrument such a volume.
Cherenkov radio signals have an attenuation length of about one kilometer. Radio is a
technology that has been successful applied in RICE.  
Acoustic signals have an even larger attenuation length of the order of $10$~km, and the ice provides a
potentially very quiet environment compared to for example the ocean.
Hybrid neutrino detectors involving acoustics and radio in addition to optics can combine the individual properties of those 
detection methods and provide a broader and more robust detection of neutrinos in a cost effective way, while it 
also allows for cross-calibration of the individual detector components for improved systematics.
Several technology options are currently actively explored.

For the season 2006-2007, prototypes for radio and acoustic sensors will be deployed along a few IceCube 
strings to study their performance.
The South Pole Acoustic Test Setup (SPATS)~\cite{Boser:2006gq} sensors, which consists of three channels
(3 piezo ceramics, spaced at $120^o$ to allow for directional sensitivity)
with low noise amplifier boards, will be deployed in three drill holes near the surface. 
The Askaryan Underice Radio Array (AURA)~\cite{dawn_arena} system consists 
of sets of 4 sensors and one transmitter for calibration purposes and test modules will be deployed
along several strings in the deep ice in the 2006-2007 deployment season.

Radio and acoustic methods are most effective for detecting electromagnetic or hadronic showers.
Detection of muon tracks by these techniques is much more difficult.
Initial sensitivity studies for EeV neutrino detection with a hybrid extension to IceCube have
been performed~\cite{Besson:2005re}.

\section{Conclusions}

IceCube's second year multistring configuration has been studied with down-going muons and artificial light sources.
The performance was found to be in good agreement with the design specifications. We demonstrated the
physics readiness of the detector and are collecting high quality physics data. 
In the coming season, we plan to more than double IceCube's current size by deploying 
another 12-14 strings.
First physics analyses have started and we expect results soon.


\end{document}